\documentclass[prb,aps,showpacs]{revtex4}
\usepackage[centertags]{amsmath}
\usepackage{amssymb}
\usepackage{amsthm}

\begin{document}

\title{Noise-induced transition in a quantum system}

\author{Pulak Kumar Ghosh, Debashis Barik and Deb~Shankar~Ray{\footnote {e-mail address:
pcdsr@mahendra.iacs.res.in}}}

\affiliation{Indian Association for the Cultivation of Science,
Jadavpur, Kolkata 700 032, India}

\begin{abstract}
We examine the noise-induced transition in a fluctuating bistable
potential of a driven quantum system in thermal equilibrium.
Making use of a Wigner canonical thermal distribution for
description of the statistical properties of the thermal bath, we
explore the generic effects of quantization like vacuum field
fluctuation and tunneling in the characteristic stationary
probability distribution functions undergoing transition from
unimodal to bimodal nature and in signal-to-noise ratio
characterizing the co-operative effect among the noise processes
and the weak periodic signal.

\end{abstract}

\maketitle

\section{Introduction}

The role of the noise in nonlinear dissipative systems has been an
intriguing issue over the last two
decades\cite{sch,lind,gar,ben,mcn,roy,luc,q,j2,a,p,aa,b,sd,hor}.
Because of their potential applications, various noise-induced
processes in zero dimensional and spatially extended systems, such
as, stochastic resonance\cite{ben,mcn,roy,luc}, coherence
resonance\cite{ark}, thermal ratchet\cite{rei,jul,mag},
noise-induced front propagation\cite{q}, noise-induced patten
formation \cite{j2,a,p,aa,b,sd} etc have been the subject of
intense investigation in widely different areas of physical,
chemical and biological sciences. An important early endeavour in
this direction is the noise-induced
 transition\cite{sch,hor} which plays a key role in understanding of the
transition of unimodular character of stationary probability
distribution to bimodal one as one varies the strength of external
 multiplicative noise in a bistable potential\cite{sch,gam,xia}.
  The theory has
been further advanced\cite{gam,xia} in several directions to
include the aspects of localization, the effect of correlation
between the noise processes, the cooperativity between the noise
processes and the signal and the associated transient
characteristics. For example, it had been shown earlier \cite{gam}
that a system in a bistable fluctuating potential driven by a very
slow modulation or static tilting in presence of an additive noise
exhibits localization in one of the wells. The stationary
probability distribution is markedly sensitive to the variation of
the correlation between the additive and multiplicative noises.
Keeping in view of these observations on noise-induced transition
in classical systems it is imperative that
quantization\cite{lin,hol,hum,juli} is likely to affect these
features since tunneling and other generic quantum effects start
playing significant role in a wider context. As prototypical
experimental systems\cite{serg} it is now possible possible to
confine electrons in semiconductors, for example, in a quantum dot
coupled to an environment via point contacts using
nanolithographic techniques to explore in detail the quantum
transport in relation to localization and tunneling.

The object of the present paper is to study noise-induced
transition in a fluctuating bistable potential in a quantum system
driven by a sinusoidal slowly varying  field in presence of
additive thermal noise of the heat bath. Our aim here is to
explore the characteristics of the stationary probability
distribution functions to understand the nature of noise-induced
transition in presence of generic quantum effects like tunneling
and vacuum fluctuations. In addition we also look for how these
changes in the stationary probability distribution functions are
reflected in the nonlinear co-operative effect between the noises
and the signal.

\section{Quantum stochastic dynamics of a driven system}

\subsection{General aspects}
To derive quantum Langevin equation from a microscopic picture we
consider the well-known standard system-reservoir model with
following form of Hamiltonian\cite{zwa}

\begin{equation}\label{2.1}
\hat{H}=\frac{\hat{p}^2}{2 m}+V(\hat{x},t)+\sum_{j=1}^N \left\{
\frac{\hat{p}^2_j}{2}+\frac{1}{2} \kappa_j (\hat{q}_j-\hat{x})^2
\right\}
\end{equation}

Here $\hat{x}$ and $\hat{p}$ are the coordinate and momentum
operators of the particle and $\{\hat{q}_j, \hat{p}_j\}$ are the
set of coordinate and momentum operators for the reservoir
oscillators coupled linearly through the coupling constants
${\kappa}_j (j=1,2,...)$. The potential $V(\hat{x},t)$ is due to
the external force field for the Brownian particle. The coordinate
and momentum operators follow the usual commutation rules
$\{\hat{x}, \hat{p}\}=i\hbar$ and $\{\hat{q}_i,
\hat{p}_j\}=i\hbar{\delta}_{ij}$. Eliminating the bath degrees of
freedom in the usual way we obtain the operator Langevin equation
for the particle

\begin{equation}\label{2.2}
m \ddot{\hat{x}}+\int^{t}_0
dt'\gamma(t-t')\dot{\hat{x}}(t')+V'(\hat{x},t) = \hat{F}(t)
\end{equation}

 where the noise operator $\hat{F}(t)$
and the memory kernel $\gamma(t)$ are given by

\begin{equation}\label{2.3}
\hat{F}(t) =
\sum_j\left[\{\hat{q}_j(0)-\hat{x}(0)\}\kappa_j\cos\omega_jt
+\kappa_j^{1/2}\hat{p}_j(0) \sin\omega_jt\right]
\end{equation}

and

\begin{equation}\label{2.4}
\gamma(t) =\sum_j\kappa_j\cos\omega_jt
\end{equation}

respectively, with $\kappa_j=\omega_j^2$. Very recently it has
been shown[28-30] that on the basis of quantum mechanical average
$ \langle...\rangle $ over the bath modes with coherent states and
the system with an arbitrary state Eq.(2.2) can be cast into the
form of the generalized quantum Langevin equation.

\begin{equation}\label{2.5}
m \ddot{x} + \int_0^{t}dt'\gamma(t-t') \dot{x}(t')+V'(x,t)= f(t) +
Q(x ,\langle\delta\hat{x}^n\rangle)
\end{equation}

where the quantum mechanical mean value of the position operator
$\langle\hat{x}\rangle =x$. Here the quantum dispersion term $Q$
to the potential, is given by

\begin{equation}\label{2.6}
Q(x ,\langle\delta\hat{x}^n\rangle) = V'(x,t) - \langle
V'(\hat{x},t)\rangle
\end{equation}

which by expressing  $ \hat{x}(t) = x(t) + \delta\hat{x}(t)$ in
$V(\hat{x},t)$ and using a Taylor series expansion around $x$ may
be rewritten as

\begin{equation}\label{2.7}
Q(x ,\langle\delta\hat{x}^n\rangle) =-\sum_{n\geq
2}\frac{1}{n!}V^{n+1}(x,t)\langle\delta\hat{x}^n\rangle
\end{equation}

The calculation of $Q$ rests on the quantum correction terms
$\langle{\delta\hat{x}^n}\rangle$ which one determines by solving
a set of quantum correction equations(as discussed in the later
part of this section). Furthermore the c-number Langevin force is
given by

\begin{equation}\label{2.8}
f(t) =
\sum_j\left[\{\langle\hat{q}_j(0)\rangle-\langle\hat{x}(0)\rangle\}
\kappa_j\cos\omega_jt +\kappa_j^{1/2}\hat{p}_j(0)
\sin\omega_jt\right]
\end{equation}

which must satisfy noise characteristics of the bath at
equilibrium ,

\begin{eqnarray}
\langle{f(t)} \rangle_S & = &
0\label{2.9}\\
\langle f(t) f(t') \rangle_S &=& \frac{1}{2} \sum_j \kappa_j\;
\hbar \omega_j \left( \coth \frac{\hbar \omega_j}{2 k T} \right)
\cos \omega_j (t-t')\label{2.10}
\end{eqnarray}

Eq.(\ref{2.10}) expresses the quantum fluctuation-dissipation
relation. The above conditions (2.9-2.10) can be fulfilled
provided the initial shifted co-ordinates
$\{\langle\hat{q}_j(0)\rangle-\langle\hat{x}(0)\rangle\}$ and
momenta $\langle{\hat{p}_j}(0)\rangle$ of the bath oscillators are
distributed according to a canonical thermal Wigner distribution
\cite{wig,hil} of the form

\begin{equation}\label{2.11}
P_j([\langle\hat{q}_j(0)\rangle-\langle\hat{x}(0)\rangle],
\langle\hat{p}_j(0)\rangle) = N \exp\left\{-\;\frac{\frac
{1}{2}\langle\hat{p}_j(0)\rangle^2 + \frac
{1}{2}\kappa_j[\langle\hat{q}_j(0)\rangle-\langle\hat{x}(0)\rangle]^2}
{\hbar\omega_j[\overline{n}(\omega_j) + \frac{1}{2}]}\right\}
\end{equation}

so that the statistical averages $\langle...\rangle_s $ over the
quantum mechanical mean value $O$ of the bath variables are
defined as

\begin{equation}\label{2.12}
\langle O_j \rangle_s=\int O_j\; P_j\; d\langle
\hat{p}_j(0)\rangle\;
d\{\langle\hat{q}_j(0)\rangle-\langle\hat{x}(0)\rangle\}
\end{equation}

Here $\overline{n}(\omega)$ is given by Bose-Einstein
distributions $(e^{\frac{\hbar\omega}{kT}}-1)^{-1}$. $P_j$ is the
exact solution of Wigner equation for harmonic oscillator
\cite{wig,hil} and forms the basis for description of the quantum
noise characteristics of the bath kept in thermal equilibrium at
temperature $T$. $N$ is the normalization constant. In the
continuum limit the fluctuation-dissipation relation (\ref{2.10})
can be written as

\begin{equation}\label{2.13}
\langle {f(t)f(t')}\rangle_s = \frac{1}{2}\;\int_0^\infty
d\omega\; \kappa(\omega)\;\rho(\omega)\; \hbar\omega\;
\coth({\frac{\hbar\omega}{2kT}})\;\cos{\omega(t-t')}
\end{equation}

where we have introduced the density of the modes $\rho(\omega)$.
Since we are interested in the Markovian limit in the present
context, we assume  $\kappa(\omega)\rho(\omega)
=\frac{2}{\pi}\gamma$, and Eq.(\ref{2.13}) then yields\cite{loi}

\begin{equation}\label{2.14}
\langle f(t)f(t')\rangle_s =2 D_q\delta(t-t')
\end{equation}

with

\begin{equation}\label{2.15}
D_q
=\frac{1}{2}\gamma\hbar\omega_0\coth{\frac{\hbar\omega_0}{2kT}}
\end{equation}

 $\omega_0$ refers to static frequency limit.
 Furthermore from Eq.(\ref{2.4}) in the continuum limit we have

\begin{equation}\label{2.16}
\gamma(t-t') = \gamma\;\delta(t-t')
\end{equation}

$\gamma$ is the dissipation constant in the Markovian limit. In
this limit Eq.(\ref{2.5}) therefore reduces to

\begin{equation}\label{2.17}
m \ddot{x} + \gamma \dot{x}+V'(x,t)= f(t ) + Q(x
,\langle\delta\hat{x}^n\rangle)
\end{equation}
In order to consider the stochastic dynamics in a bistable system
with fluctuating barrier height we consider the potential as,
\begin{equation}\label{2.18}
V(x,t) =
-\left[\frac{a}{2}+\frac{\xi(t)}{2}\right]x^2+\frac{b}{4}x^4
\end{equation}

$\xi(t)$ is the external fluctuations with zero mean and is
characterized by the following equations
\begin{equation}\label{2.19}
\overline{\xi(t)} =0\;\;; \;\;\;\;\;\;\; \overline{\xi(t)\xi(t')}
= 2Q_I \delta(t-t')
\end{equation}
where $Q_I$ is the strength of noise. Care must be taken to
distinguish between the quantum mechanical mean
$\langle{\hat{O}}\rangle$ $(=O)$, the statistical average over
 the quantum mechanical mean $\langle{O}\rangle_s$ denoting thermal
 bath average and the statistical average over the external noise
 $\overline{O}$.

Attention is restricted here to overdamped Langevin equation in
one variable $x$ so that we write

\begin{equation}\label{2.20}
\gamma \dot{x}= (ax-bx^3) + x\xi(t)+f(t ) + Q(x
,\langle\delta\hat{x}^n\rangle)
\end{equation}
$Q(x,\langle\delta\hat{x}^n\rangle)$ being the correction term due
to nonlinearity of the potential is too small to destroy
bistablity or the symmetry of the system. If we now apply an weak
periodic forcing to the particle, the double-well potential is
tilted asymmetrically up and down and the corresponding stochastic
differential equation reads as
\begin{equation}\label{2.21}
\gamma \dot{x}= (ax-bx^3) + x\xi(t)+f(t ) + Q(x
,\langle\delta\hat{x}^n\rangle)+A\cos{\omega t}
\end{equation}
 The Fokker-Planck equation corresponding to Eq.(\ref {2.21}) is given
by

\begin{equation}\label{2.22}
\frac{\partial P(x,t)}{\partial t}=\frac{\partial}{\partial
x}\left[B(x,t)+\frac{\partial}{\partial x}D(x)\right]P(x,t) \;,
\end{equation}
where
\begin{equation}\label{2.23}
B(x,t)=-\frac{1}{\gamma}\left[ (ax-bx^3) + \frac{xQ_I}{\gamma} -
Q(x ,\langle\delta\hat{x}^n\rangle)+A\cos{\omega t} \right]\;,
\end{equation}
and,
\begin{equation}\label{2.24}
D(x)=\frac{1}{\gamma^2}\left[Q_Ix^2+D_q \right]\;.
\end{equation}
Eq.(\ref {2.22}) can then be rearranged into the following form
\begin{equation}\label{2.25}
\frac{\partial P(x,t)}{\partial
t}=\frac{1}{\gamma}\frac{\partial}{\partial
x}\left[-ax+bx^3-\frac{Q_I x}{\gamma}-A\cos{\omega
t}-Q(x,\langle\delta\hat{x}^n\rangle)+\frac{1}{\gamma}\frac{\partial}{\partial
x}(D_q+Q_I x^2)\right]P(x,t)
\end{equation}
The quantum nature of the stochastic dynamics is manifested in two
terms; the quantum correction to the classical potential $Q$ and
the quantum diffusion coefficient $D_q$ characterizing the thermal
bath. Eq.(\ref {2.25}) is the quantum counterpart of Eq.(5)
derived in Ref \cite{gam}. Furthermore Eq.(\ref {2.25}) formally
takes care of quantum correction to all orders and the vacuum
field effect of the bath at zero temperature.

\subsection{Static tilting}
In the presence of static tilting, i.e., $\omega=0$, the
stationary solution of the Fokker-Planck equation reads as

\begin{equation}\label{2.26}
P_0(x,A)=N_0{\left[
x^2+\frac{D_q}{Q_I}\right]}^{-\frac{1}{2}+\gamma k\left(1+\frac{k
D_q}{2\Delta V}\right)}\;\;G(x)\;\;\exp{\left[
-\gamma\left(\frac{kx^2}{x_0^2}-\frac{A}{|x|}\right)\right]}
\end{equation}
The term $G(x)$, arises due to quantum corrections to the
potential and is given by the following expression
\begin{equation}\label{2.27}
G(x)=\exp{\left[ -\int_0^x\;\;dx'\;\;\; \frac{\gamma \;\;
Q(x',\langle\delta\hat{x}^n\rangle)}{Q_I\left(x'^2+\frac{D_q}{Q_I}\right)}
\right]}
\end{equation}
$N_0$ is a suitable normalization constant, $k=\frac{a}{2Q_I}$;
$\Delta V$ and $\pm x_0$ denote the barrier height and stable
minima of the bistable potential $-\frac{a}{2}x^2+\frac{b}{4}x^4$,
respectively, given by the following expressions
\begin{equation}\label{2.28}
\Delta V = \frac{a^2}{4b}\;\;;\;\;\;\;\;x_0 =
\left(\frac{a}{b}\right)^{\frac{1}{2}}
\end{equation}
To proceed further it is necessary to find out the quantum
correction term $Q(x ,\langle\delta\hat{x}^n\rangle)$ more
explicitly. From Eq.(\ref{2.7}) we have

\begin{eqnarray}\label{2.29}
Q(x ,\langle\delta\hat{x}^n\rangle) =
-\frac{V'''(x,t)}{2!}\langle\delta\hat{x}^2\rangle
-\frac{V''''(x,t)}{3!}\langle\delta\hat{x}^3\rangle+....
\end{eqnarray}
For a bistable potential $V^{n+1}(x,t)=0$ for $n\geq 4$. For a
minimum uncertainty state $\langle\delta\hat{x}^2\rangle=O(\hbar)$
and therefore the higher order terms of the order
$\langle\delta\hat{x}^3\rangle$ can be safely neglected.
Furthermore since the stochastic system is overdamped the quantum
corrections are primarily relevant in the time scale
$O$($\frac{1}{\gamma}$) It is therefore important to calculate the
leading order quantum correction
$-\frac{V'''(x,t)}{2}\langle\delta\hat{x}^2\rangle$.
  To this end we consider
the overdamped version of operator equation (\ref{2.2}) and use
$\hat{x}( t)=x( t)+ \delta\hat{x}(t)$, where $x(
t)=\langle\hat{x}(t)\rangle$  is the quantum mechanical mean value
of the operator $\hat{x}$. By construction
$[\delta\hat{x},\delta\hat{p}]=i\hbar$ and $\langle\delta
\hat{x}\rangle=0$. We then make use of the overdamped version
Eq.(\ref{2.17}) and obtain the operator equation after quantum
mechanical average over bath modes with product separable coherent
states

\begin{equation}\label{2.30}
\gamma\delta\dot{\hat{x}} +V''(x,t)\delta\hat{x}+ \sum_{n\geq
2}\frac{1}{n!}{V}^{n+1}(x,t)(\delta\hat{x}^n- \langle
\delta\hat{x}^n\rangle)=0
\end{equation}

With the help of operator equation (\ref{2.30}) we obtain the
coupled equations for $\langle\delta\hat{x}^n\rangle$

\begin{equation}\label{2.31}
\frac{d}{dt}\langle\delta\hat{x}^2\rangle=-\frac{1}{\gamma}\left[2V''(x,t)\langle\delta\hat{x}^2\rangle
+V'''(x,t)\langle\delta\hat{x}^3\rangle\right]
\end{equation}

\begin{equation}\label{2.32}
\frac{d}{dt}\langle\delta\hat{x}^3\rangle=-\frac{1}{\gamma}\left[
3V''(x,t)\langle\delta\hat{x}^3\rangle
+\frac{3}{2}V'''(x,t)\langle\delta\hat{x}^4\rangle-\frac{3}{2}V'''(x,t)
\langle\delta\hat{x}^2\rangle^2\right]
\end{equation}
and so on. To take into account of the leading order contribution
$\langle\delta\hat{x}^2\rangle$ to $Q(x
,\langle\delta\hat{x}^n\rangle)$ explicitly we retain only the
first term in Eq.(\ref {2.31})
\begin{equation}\label{2.33}
d\langle\delta\hat{x}^2\rangle =
-\frac{2}{\gamma}V''(x,t)\langle\delta\hat{x}^2\rangle dt
\end{equation}
The overdamped deterministic motion on the other hand gives
$\gamma dx = -V'(x,t)dt$ which can be used in Eq.(\ref{2.33}) to
eliminate $dt$ and obtain after integration
\begin{equation}\label{2.34}
\int^{\langle\delta\hat{x}^2\rangle}_{\langle\delta\hat{x}^2\rangle_c}
\frac{d\langle\delta\hat{x}^2\rangle}{\langle\delta\hat{x}^2\rangle}=2\int^{x}_{x_c}
\frac{V''(x,t)}{ V'(x,t)}
\end{equation}
where $x_c$ is a quantum mechanical mean position at which
$\langle\delta\hat{x}^2\rangle$ becomes minimum, i. e.,
$\langle\delta\hat{x}^2\rangle_{x_c}=\frac{1}{2}\hbar /\omega_0$,
$w_0$ being defined in Eq.(\ref{2.15}). Explicit integration
yields

\begin{equation}\label{2.35}
\langle\delta\hat{x}^2\rangle = \Delta_c\left[ V'(x,t)\right]^2
\end{equation}
where $\Delta_c$ is  given by
$\Delta_c=\frac{\langle\delta\hat{x}^2\rangle_{x_c}}{\left[
V'(x_c,t)\right]^2}$. This quantum correction parameter can be
estimated as follows:

From Eq.(\ref{2.18}) we have
\begin{equation}\label{2.36}
\left[V'(x,t)\right]^2 = x^2
\left[a^2+\xi^2(t)+2a\xi(t)+2(a+\xi(t))bx+b^2x^4\right]
\end{equation}
By averaging Eq.(\ref{2.36}) over external noise $\xi(t)$ using
Eq.(\ref{2.19}) we have from Eq.(\ref{2.35})
\begin{equation}\label{2.37}
\langle\delta\hat{x}^2\rangle = \Delta_cx^2
\left(a^2+Q_I+b^2x^4-2abx^2\right)
\end{equation}
The minimum of $\langle\delta\hat{x}^2\rangle$ at $x=x_c$ can be
obtained from Eq.(\ref{2.37}) as
$x_c=\left[\frac{a^2+Q_I}{4ab}\right]^{1/2}$ by neglecting $b^2$
term (since $a\gg b$).

 With the help of Eq.(\ref{2.35}) it is easy to obtain from Eq.(\ref {2.29}) the
leading order quantum correction averaged over external noise as
given by (we have dropped the overbar sign used in Eq.(\ref
{2.19}), for convenance)
\begin{equation}\label{2.38}
Q(x,\langle\delta\hat{x}^n\rangle) = -6\Delta_qbx^3
\left(a^2+Q_I+b^2x^4-2abx^2\right)
\end{equation}
where $\Delta_q=\frac{\Delta_c}{2}$.
$Q(x,\langle\delta\hat{x}^n\rangle)$ in Eq.(\ref{2.38}) is an odd
function, which contributes to the potential force term, so that
the system retains its symmetry (i.e., reflection symmetry of the
bistable potential remain unchanged)

Now we are in position to discuss the properties of the stochastic
processes $x(t)$, in terms of its steady state distribution
function $P(x,A)$ for different regimes of coherent and inherent
system parameters.

In absence of the static tilting i.e., $A=0$, the process $x(t)$
diffuses on the entire $x$ axis, the relevant distribution
function is symmetric for $x\rightarrow -x$ and
$\langle{x(t)}\rangle=0$. As temperature of the system decreases
the probability distribution decreases at the origin, since $D_q$
decreases with temperature. In the classical limit as
$T\rightarrow 0$, $D_q\rightarrow \gamma k_B T \rightarrow 0 $ we
observe $ P(x,0)\rightarrow 0$ at the origin. This is shown in
solid line in Fig.$1$. So the process $x(t)$ remains confined in
any one half axis depending on the initial condition. Unlike the
classical limit at $T=0$ the distribution function $P(x,0)$ in the
quantum limit takes the following form;
\begin{equation}\label{2.39}
P_0(x,0)=N_0{\left[ x^2+\frac{D_q(0)}{Q_I}\right]}^{-\frac{1}{2}+
k\left(1+\frac{k D_q(0)}{2\Delta V}\right)}\exp{\left[
-\frac{kx^2}{x_0^2}\left(1-\Delta_1\right)\right]}
\end{equation}

Here $\gamma$ is assumed to be unity, $D_q(0)$ is zero point
diffusion coefficient having non-zero value
$\frac{1}{2}\hbar\omega_0$, $\Delta_1$ is the contribution due to
system nonlinearity correction as given by
\begin{equation}\label{2.40}
\Delta_1=6\Delta_q\left(a^2+Q_I\right)\;.
\end{equation}
Hence the process $x(t)$ does not remain confined in the one half
axis depending on the initial condition even at zero temperature
due to quantum mechanical tunneling. This is an important aspect
of bath quantization. The distribution function peaks at $x_m\sim
\pm {\left(
\frac{nx_0^2}{k\left(1-\Delta_1\right)}-\frac{D_q(0)}{Q_I}\right)}^
{\frac{1}{2}}$, where  $n={-\frac{1}{2}+k\left(1+\frac{k
D_q(0)}{2\Delta V}\right)}$. Thus compared to classical limit the
peaks of the distribution function shift slightly towards the
origin due to zero point contribution of the bath oscillators. All
the above observations are in the limit $a>Q_I$ and are shown in
Fig.1 and Fig.2. In the zero temperature limit as the external
noise intensity is increased to $Q_I=a$, the distribution function
Eq.(\ref{2.39}) attains an unimodal form since
$\frac{D_q(0)}{\Delta V}\leq 1$; a phenomenon like phase
transition occurs both in the classical and the quantum limit.
This is presented in the Fig.3. In the quantum case the
distribution function deviates from the Gaussian shape due to
quantum corrections arising out of higher order nonlinearity. It
is important to note that the quantum particle is more localized
than the classical particle over the $x$-axis, a feature typical
of quantum localization as shown in this case. For $Q_I>a$, the
distribution function Eq.(\ref{2.39}) tends to be singular at the
origin since $n<0$ and the appearance of long tails is
characterized by stochastic stabilization.

\subsection{Periodic tilting}
When the periodic tilting is present $\omega\neq 0$ and $A\neq 0$,
the process $x(t)$ is no-longer stationary and the time-dependent
distribution function $P(x,A,t)$ is necessary to describe its
steady-state. However in the limit of low forcing
frequency\cite{gam} there is enough time for the system to reach
the local equilibrium during the period of $\frac{1}{\omega}$.
Then the quasi-steady state distribution function reads as
\begin{equation}\label{2.41}
P(x)=NG(x){\left[ x^2+\frac{D_q}{Q_I}\right]}^{-\frac{1}{2}+
k\left(1+\frac{k D_q}{2\Delta V}\right)}\;\;\exp{\left[
-\frac{kx^2}{x_0^2}-L(x)\right]}
\end{equation}
where the time $\tau$ is used as fixed parameter so that
\begin{equation}\label{2.42}
L(x)=\frac{A\cos\omega \tau}{\left({D_q Q_I}\right)^{\frac{1}{2}}}
\arctan\left[\left(\frac{Q_I}{D_q}\right)^{\frac{1}{2}}x\right]
\end{equation}
The quasi-steady state distribution function $P(x)$ is plotted in
Fig.4 and Fig.5 as a function of position coordinate when the
strength of additive and multiplicative noises $D_q $ and $Q_I$,
respectively, are varied. The probability density function $P(x)$
is depicted in Fig.4 for several values of quantum diffusion
constant and for a fixed value of $A$(=-0.1). The peak located at
$x=-(\frac{a}{b})^{\frac{1}{2}}$ is much higher than that at
$x=+(\frac{a}{b})^{\frac{1}{2}}$ when $D_q=0.9$. When $D_q$ is
increased to $10$, the distribution function $P(x)$ gets broadened
and the left peak is decreased to almost equal height to the right
one. Fig.5 depicts a situation where the multiplicative noise
strength is varied. It is observed that the peak of the
distribution function at $x=-(\frac{a}{b})^{\frac{1}{2}}$ is
decreased and moves towards the origin when the strength of the
external noise is increased from $0.08$ to $0.7$ with a gradual
lose of Gaussian form.

\section{The co-operative effect; Multiplicative Stochastic resonance }
In the previous section we have been analyzed the stochastic
process $x(t)$ in terms of the probability distribution functions.
We have compared the distribution function of quantum mechanical
mean values with the corresponding classical one in the different
parameter regime which reflect the profound effect of bath
quantization at very low temperature. But this is not sufficient
to describe completely the cooperative effect among the inherent
thermal noise, the external noise and weak periodic signal. To
analyze such nonlinear cooperative effect, i.e., the phenomenon of
stochastic resonance in the bistable system the most common way to
quantify the effect is through signal-to-noise ratio (SNR). SNR is
defined as the ratio of the strength of $\delta$-spike of the
power spectrum to the back ground noise level. Following
\cite{mcn,luc} the expression of SNR in the bistable system can be
derived from the well-known two-state model approach as
\begin{equation}\label{3.1}
  SNR=\frac{\pi}{2}W_0\left(\frac{Ax_0}{D_q}\right)^2\left(1-\frac{1}{2}
  \left(\frac{Ax_0}{D_q}\right)^2
  \left[\frac{4W_0^2}{4W_0^2+\omega^2}\right]\right)^{-1}
\end{equation}
where $W_0$ is the Kramer's hopping rate over the activated
barrier in absence of periodic force and $\pm x_0$ are the two
stable states.

The simplest way to calculate Kramer's rate is through mean first
passage time (MFPT). The exact expression for MPFT \cite{str} for
a particle to reach the final point $x_2$, the quantum mechanical
mean position from an initial point $x_1$ is given by
($W_0=\frac{1}{T_m}$)
\begin{equation}\label{3.2}
T_m(x_1\rightarrow x_2)=\int^{x_2}_{x_1}
\frac{dx}{D(x)P_s(x)}\int_{-\infty}^x P_s(x)dy
\end{equation}
From Eq.(\ref{2.22} - \ref{2.24}), we obtain the stationary
distribution $P(x)$ ($A=0$ and $\gamma=1$) as
\begin{equation}\label{3.3}
P_s(x)=ND^{-\frac{1}{2}}(x)
\exp{\left[\frac{-\overline{U}(x)}{Q_I}-\int_0^x\frac{Q(x,\langle\delta\hat{x}^n\rangle)}{D(x)}dx\right]}
\end{equation}
where
\begin{equation}\label{3.4}
D(x)=Q_Ix^2+D_q\;\;;\;\;\;\;\overline{U}(x)=-\int_0^x\frac{ay-by^3}
{\left(y^2+\frac{D_q}{Q_I}\right)dy}
\end{equation}
The two points $x_1$ and $x_2$ are $-x_0$ and $0$, respectively.
Now using steepest-descent approximation we obtain the expression
for $T_m$ in the usual way
\begin{equation}\label{3.5}
T_m(x_1\rightarrow x_2)=2\pi
|U''(x_1)U''(x_2)|^{-\frac{1}{2}}\exp{\left[\frac{\overline{U}(x_1)-\overline{U}(x_2)}
{Q_I}\right]} \exp{\left[\phi(x_2)-\phi(x_1)\right]}
\end{equation}
where
\begin{equation}\label{3.6}
\phi(x)=\int_0^x\frac{Q(y,\langle\delta\hat{y}^n\rangle)}{D(y)}dy\;\;;\;\;\;\;\;
U(x)=-\frac{a}{2}x^2+\frac{b}{4}x^4
\end{equation}
Having determined the hopping rate we are in position to explain
the effect of different parameters on the signal-to-noise ratio.
From the equation the effect of increasing input signal $A$ is
quite straightforward. On increasing the input signal one observes
the increase of SNR. The implication of varying the signal
frequency is, however, complicated. To explore this effect
 we present in Fig.6\ the variation of signal-to-noise ratio as
function of temperature for several values of signal
frequency($\omega$). The solid curve represents the value of SNR,
for lowest frequency($\omega=0.0001$) and dashed refers to the
same for the largest frequency($\omega=0.1$). The SNR decreases as
the frequency is increased in the lower temperature range. When
the signal frequency is low, the particle gets time to approach a
local equilibrium, and it has a high likelihood of hopping from
upper well to the lower well during the half cycle. Hence a type
of resonance phenomenon is expected to occur at optimum level of
thermal fluctuation. At higher temperature such type of nonlinear
cooperative effect is destroyed due to substantial probability of
hopping back to the upper well.
 At the high frequency regime on the other hand such nonlinear
  cooperative effect
 is less likely to be observed, because a fewer number of particles
 find the time to hop to the lower well during each cycle. On
 increasing temperature one may increase the inter-well transition
 rate, but at the risk of being kicked in the anti-phase
 direction. Although SNR passes through a maximum as one varies
 the thermal noise strength it may be considered as
 some type of compromise between input signal and inherent thermal
 fluctuation rather than a resonance.

 Fig.7 depicts the variation of SNR as a function of temperature
comparing the classical and the quantum limits and shows that SNR
is larger in quantum limit at low temperature and the difference
become insignificant at high temperature as expected.
 It is also interesting to note that in contrast to
 classical limit SNR is not zero at absolute zero due to zero
 point contribution of the thermal bath.

\section{conclusion}
Based on the study of stationary probability distribution function
and signal-to-noise ratio we have investigated the problem of
quantum stochastic dynamics of a system in a bistable fluctuating
potential which is driven by a slow periodic or static field. Our
conclusions are summarized as follows:

(i) Temperature profoundly affects the characteristic of the
bimodal stationary probability distribution function which reveals
that in contrast to classical case, where the system remains
confined in one of the wells, quantization allows a finite
population at barrier top even at zero due to vacuum field effect
of the thermal bath in absence of tilting and for a low noise
strength of the modulation for fluctuating well. Thus although at
zero temperature the passage between the wells is classically
forbidden, quantum tunneling makes the passage smooth for
population transfer.

(ii) For a static tilting and a higher strength of stochastic
modulation of the well, the unimodal stationary probability
distribution becomes
 more localized and acquires non-Gaussian character on
 quantization as compared to that for the corresponding classical system.

 (iii) Very slow external periodic driving may localize the system
  in one of the wells. The peak of the distribution function
  shifts towards the origin as the strength of modulation of
  fluctuating potential is increased.

(iv) The effect of quantization also makes its presence felt
significantly in the signal-to-noise ratio at low temperature.

 The
method presented here takes care of the statistical properties of
the thermal bath in terms of Wigner canonical thermal distribution
which remains a valid  pure state distribution even at absolute
zero. This help us to consider the vacuum field limit in the
treatment of stochastic processes the governing equation of which
is classical in form but quantum mechanical in its content.
Moreover although formally one takes into account of the quantum
correction terms due to nonlinearity of the potential to all
orders in the stationary distribution function, it is possible to
estimate these corrections order by order for actual calculation
effectively to a good degree of accuracy. We thus hope that our
observations will be useful for experimental studies on tailored
quantum systems like quantum dots at very low temperature.

\acknowledgments Thanks are due to the Council of Scientific and
Industrial Research, Government of India, for a fellowship (PKG).

\newpage

\begin{center}
{\bf Figure Captions}
\end{center}

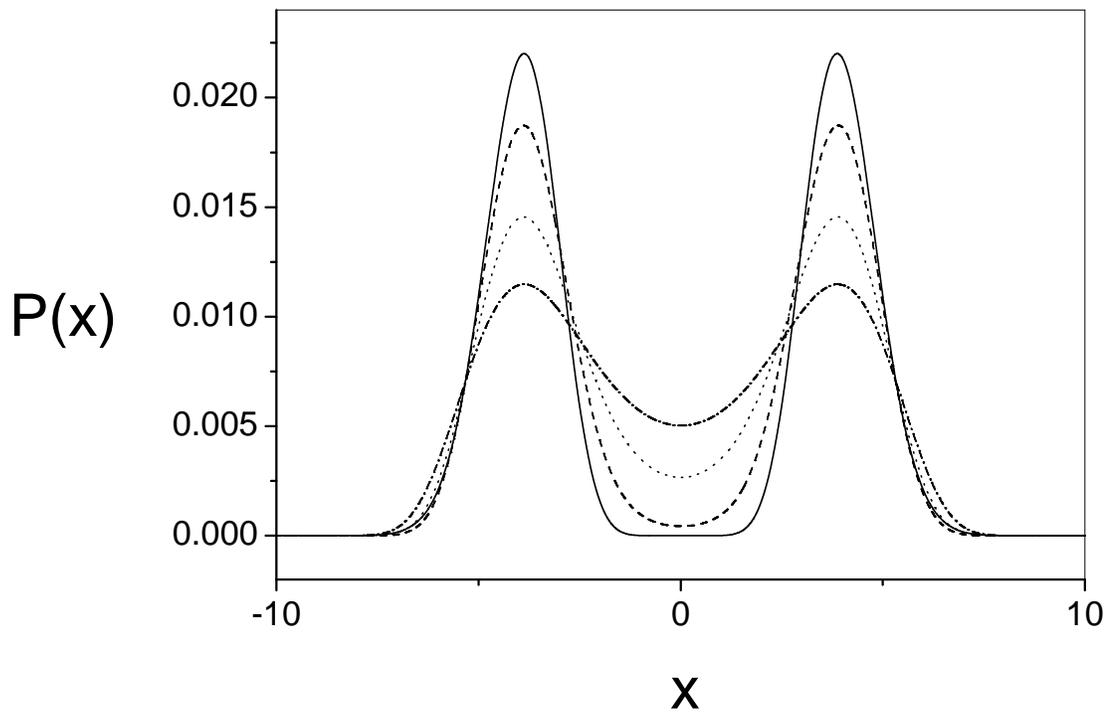
Fig.1: Variation of probability density function ($P(x)$) as a
function of quantum mechanical mean position(x) in absence of
tilting (A=0) for the parameter set $a=1.0$, $b=0.06$,
$\Delta_q=0.02$, $Q_I=0.1$ and (i) $D_q=0.0$($\gamma k_BT=0$) and
$\Delta_q=0.0$ (solid line) [classical] (ii) $D_q=0.5$ (dashed
line), (iii) $D_q=3.0$ (dotted line) and (iv) $D_q=5.00$ (dash-dot
line).

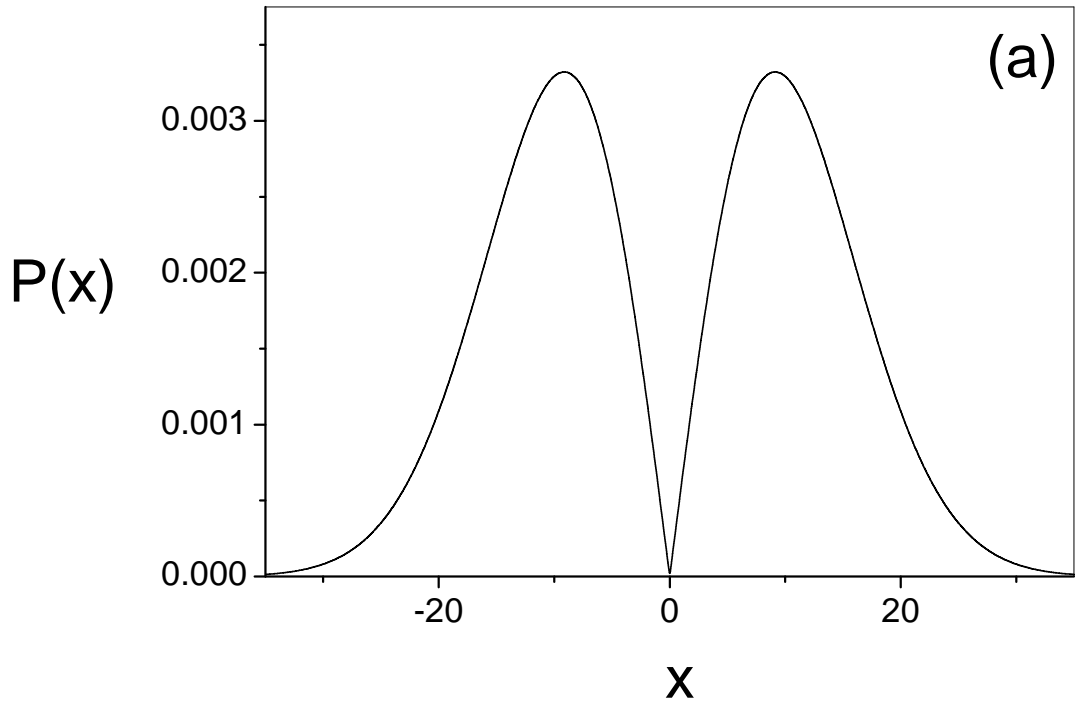
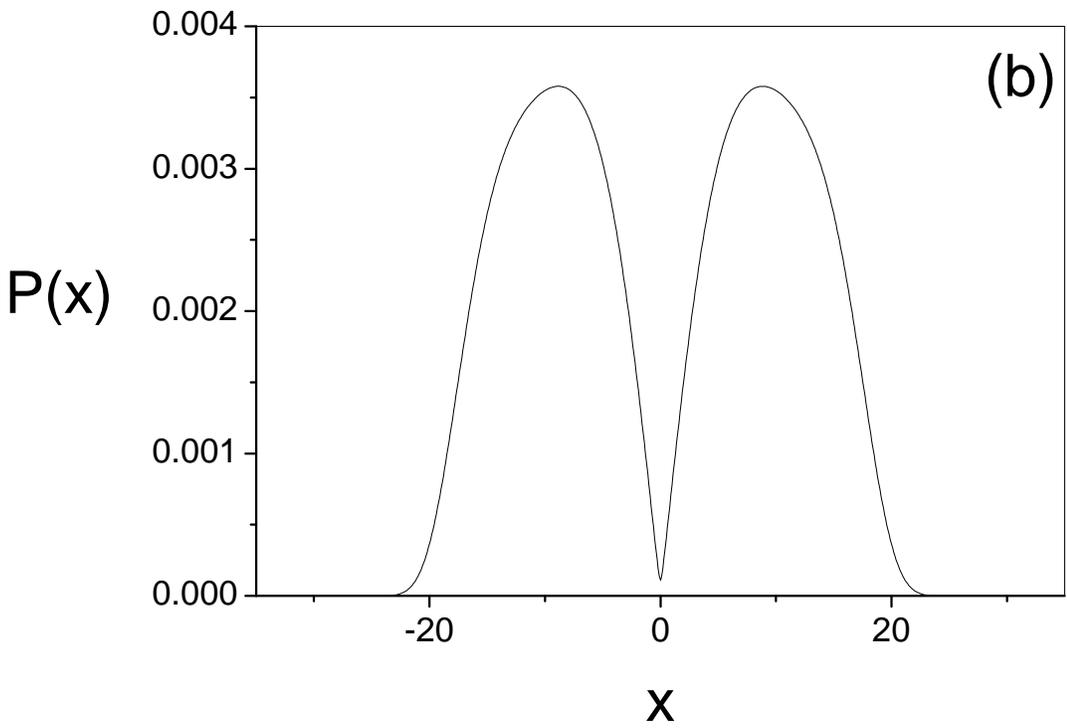
Fig.2: Variation of P(x) as a function of x at zero temperature
comparing quantum(b) and classical(a) limits for the parameter set
$a=1.0$, $b=0.006$, $Q_I=0.5$, $A=0.0$ and $\Delta_q=0.02$.

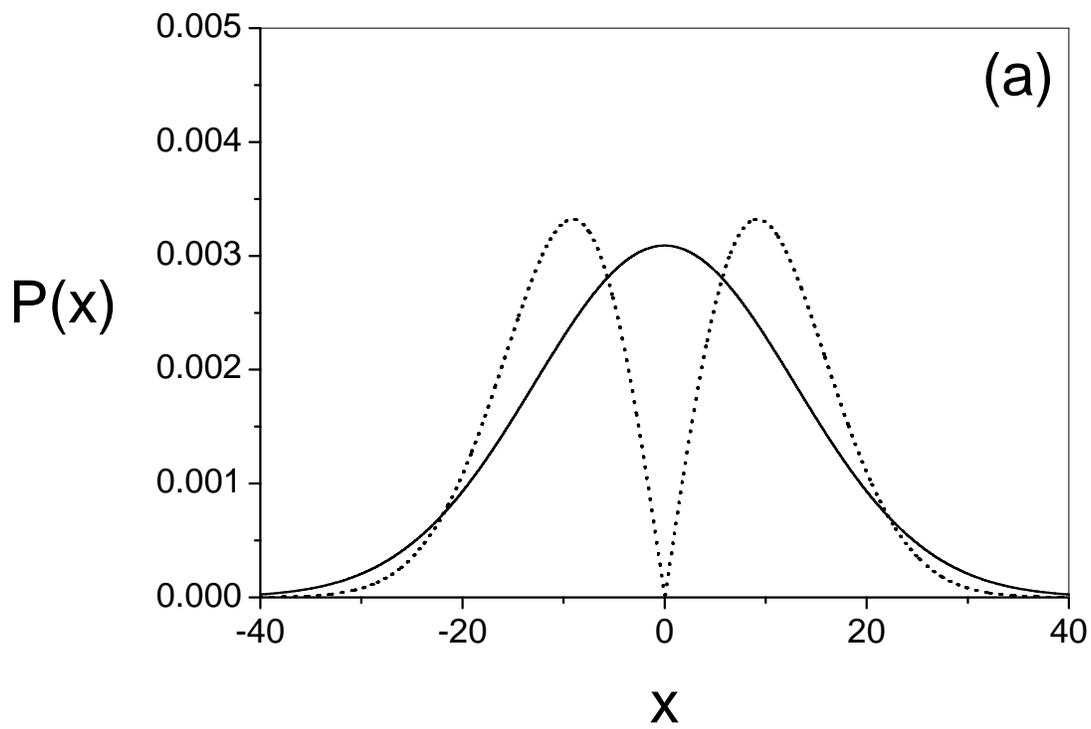
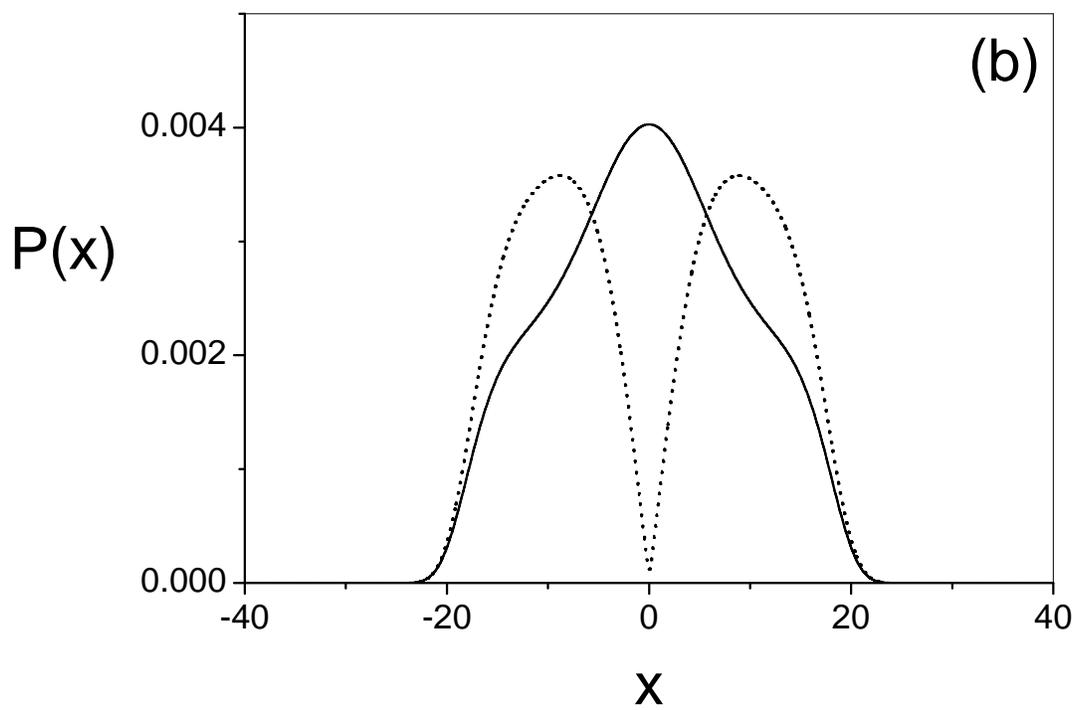
Fig.3: Variation of P(x) as a function of x at zero temperature
limit comparing quantum(b) and classical(a) phase transition for
the parameter set $a=1.0$, $b=0.006$, $A=0.0$, $\Delta_q=0.02$ and
(i) $a>Q_I=0.5$ (dotted line) and $Q_I=a=1.0$ (solid line)

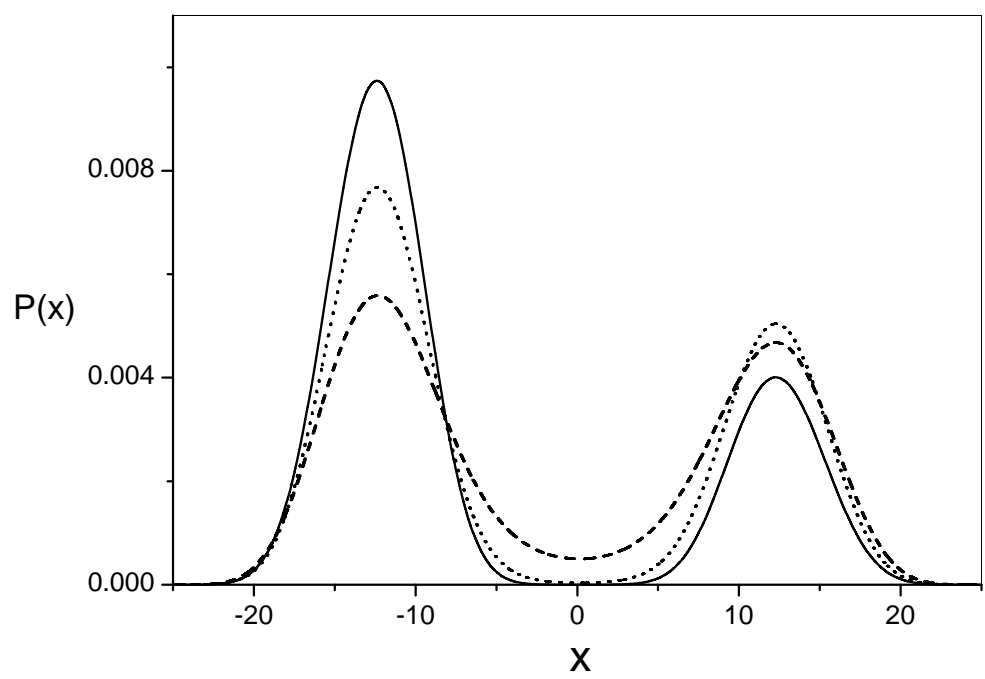
Fig.4: Variation of $P(x)$ with  $x$  for different values of
 quantum quantum diffusion coefficient ($D_q$), $D_q=0.9$ (solid
line), $D_q=3.5$(dotted line) and $D_q=10.0$ (dashed line) in
presence of tilting (A=-0.1) for $a=1.0$, $b=0.006$, $Q_I=0.1$,
$\omega =0.1$, $\Delta_q=0.02$ and $\tau=1 $

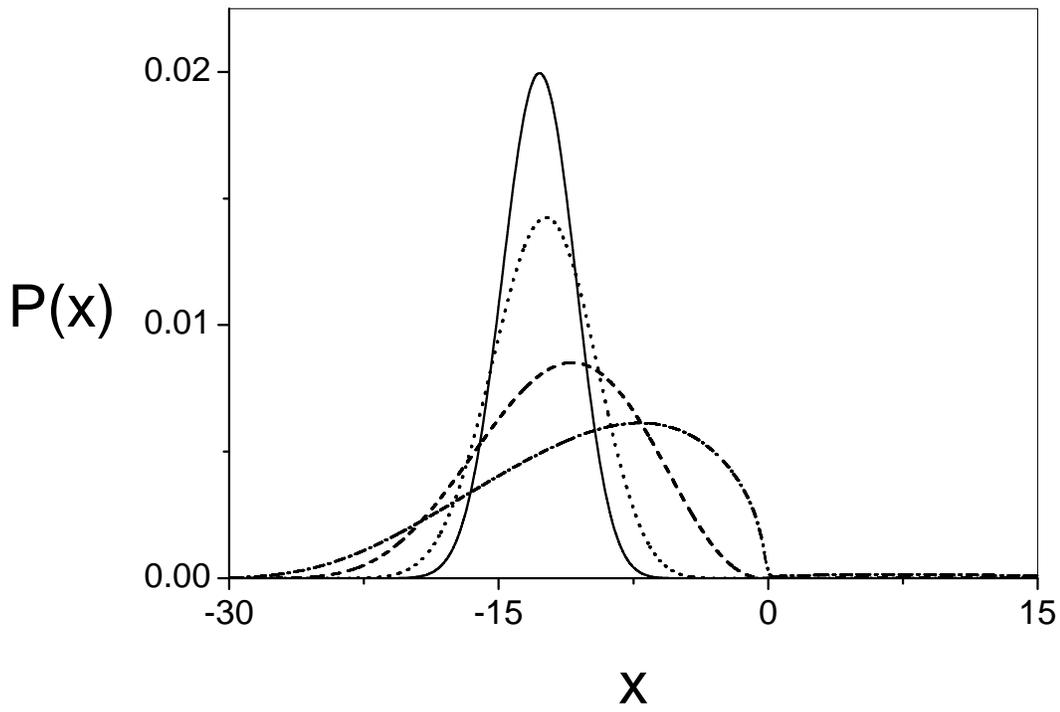
Fig.5: Variation of $P(x)$ with  $x$  for different values of
 external multiplicative noise strength ($Q_I$), $Q_=0.05$ (solid
line), $Q_I=0.1$(dotted line), $Q_I=0.3$ (dashed line) and
$Q_I=0.7$ (dash-dot line) in presence of tilting (A=-0.1) for
$a=1.0$, $b=0.006$, $Q_I=0.1$, $\omega =0.1$, $\Delta_q=0.02$ and
$\tau=1 $

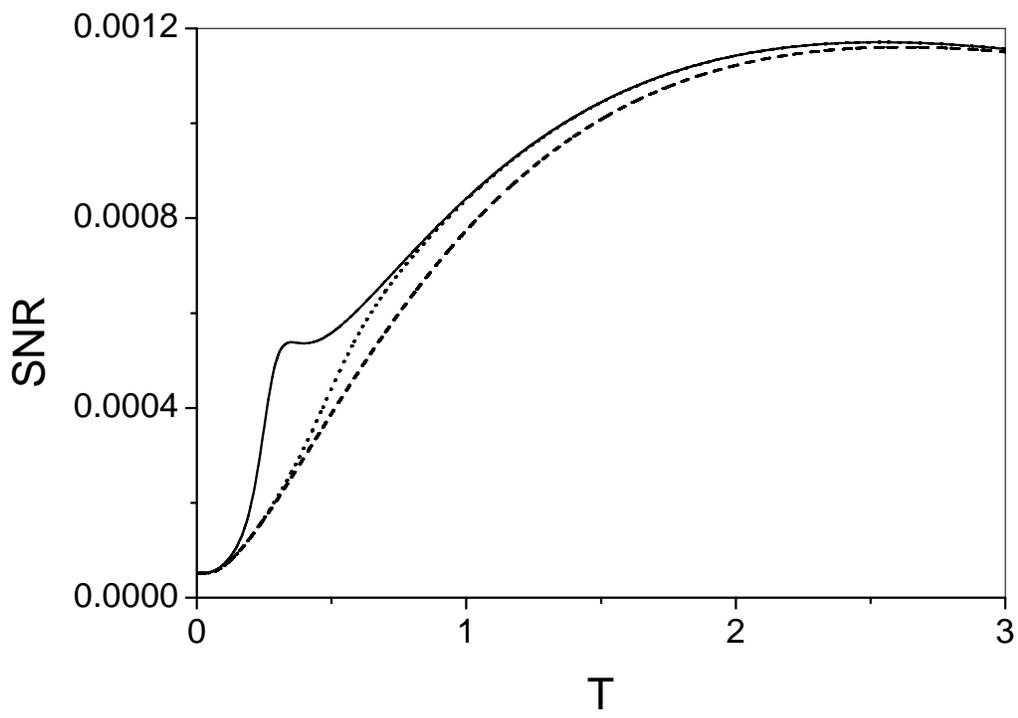
Fig.6: Variation of Signal-to-noise(SNR) ratio with temperature(T)
for the parameter set $a=2.5$, $b=0.001$, $Q_I=3.0$,
$\Delta_q=0.02$, $A=0.008$ and (i) $\omega=0.0001$ (solid line),
(ii) $\omega=0.001$ (dotted line) and (iii) $\omega=0.1$ (dashed
line)

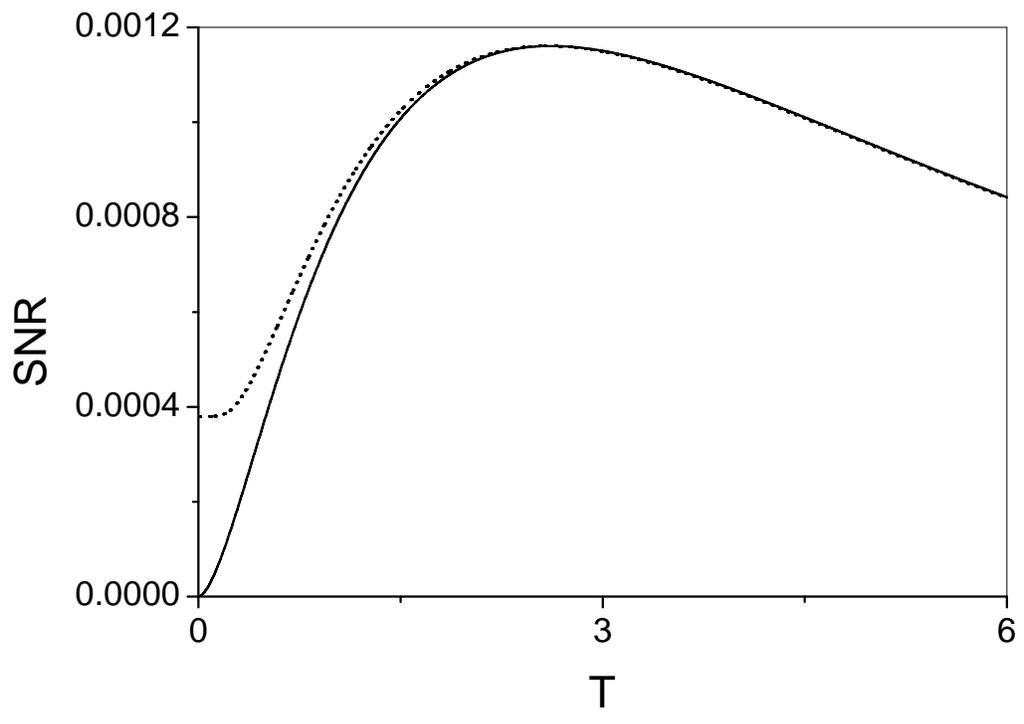
Fig.7: A plot of signal-to-noise ratio vs temperature comparing
quantum and classical limit for the  parameter set $a=2.5$,
$b=0.001$, $Q_I=3.0$, $\Delta_q=0.01$, $A=0.008$  and $\omega=0.1$
(i) classical case (solid line), (ii) quantum case (dotted line)

\end{document}